\documentclass[fleqn,10pt]{wlscirep}

\usepackage{rotating}
\usepackage{graphicx}
\usepackage{color}
\usepackage{booktabs}
\usepackage{tikz}
\usetikzlibrary{arrows}
\usetikzlibrary{snakes}

\newif\iffigsinpdf
\title{Aligning 415\,519 proteins in less than two hours on~PC}

\author[1,*]{Sebastian Deorowicz}
\author[1]{Agnieszka Debudaj-Grabysz}
\author[1]{Adam Gudy\'s}
\affil[1]{Institute of Informatics, Silesian University of Technology, Akademicka 16, 44-100 Gliwice, Poland}
\affil[*]{sebastian.deorowicz@polsl.pl}


\begin{abstract}
Rapid development of modern sequencing platforms enabled an unprecedented growth of protein families databases.
The abundance of sets composed of hundreds of thousands sequences is a great challenge for multiple sequence alignment algorithms.
In the article we introduce FAMSA, a new progressive algorithm designed for fast and accurate alignment of thousands of protein sequences.
Its features include the utilisation of longest common subsequence measure for determining pairwise similarities, a novel method of gap costs evaluation, and a new iterative refinement scheme.
Importantly, its implementation is highly optimised and parallelised to make the most of modern computer platforms.
Thanks to the above, quality indicators, namely sum-of-pairs and total-column scores, show FAMSA to be superior to competing algorithms like Clustal Omega or MAFFT for datasets exceeding a few thousand of sequences. The quality does not compromise time and memory requirements which are an order of magnitude lower than that of existing solutions. 
For example, a family of 415\,519 sequences was analysed in less than two hours and required only 8GB of RAM.

FAMSA is freely available at \url{http://sun.aei.polsl.pl/REFRESH/famsa}.
\end{abstract}

\begin{document}
\flushbottom
\maketitle
\thispagestyle{empty}

\section*{Introduction}
The multiple sequence alignment (MSA) is one of the most important analyses in molecular biology. Majority of algorithms for the MSA problem conform to a progressive heuristics~\cite{CMCKBEN2015}. The scheme includes three stages: (I) calculation of a similarity matrix for investigated sequences, (II) a guide tree construction, (III) greedy alignment according to the order given by the tree. Pairwise similarities can be established variously. Some algorithms use accurate, but time-consuming methods like calculating pairwise alignments of highest probability~\cite{Tho1994} or maximum expected accuracy~\cite{Do2005}. Others employ approximated, though faster approaches, e.g., tuple matching~\cite{Notredame2000, E2004}. As sizes of protein families to be analysed has been constantly increased, the necessity to calculate all pairwise similarities has become a bottleneck of alignment algorithms. Therefore, many attempts have been made to accelerate this stage. Kalign~\cite{LS2005} and Kalign2~\cite{LFS2009} employ for similarity measurement, respectively, Wu-Manber~\cite{Wu1992} and Muth-Manber~\cite{Mut1996} fast string matching algorithms. This allows thousands of sequences to be aligned in a reasonable time. The idea has been further extended by the authors of the presented research in Kalign-LCS~\cite{DDG2014}, which introduced to Kalign2 pipeline longest common subsequence for similarity measurement. This improved both, alignment quality as well as execution time. Nevertheless, the recent developments in high throughput sequencing confront biologists with the necessity to align protein families containing tens of thousands of members~\cite{}. Progressive algorithms which calculate and store all pairwise similarity distances, were inapplicable for problems of such sizes due to excessive time and memory requirements.

An introduction of PartTree, a divisive sequence clustering algorithm for building a guide tree without calculating all pairwise similarities~\cite{Kat2007}, was one of the ideas to tackle the problem. With average time complexity of $O(k\log k)$ and space complexity of $O(k)$ ($k$ is the number of sequences in the input set), PartTree was successfully adopted by MAFFT~6 package~\cite{Katoh2008} allowing tens of thousands of sequences to be aligned on a typical desktop computer. A different approach was presented in Clustal Omega~\cite{SWD2011}. It uses mBed, an algorithm for embedding sequences into a lower-dimensional space~\cite{Bla2010}, which requires only $O(k\log k)$ exact similarity values to approximate others. The embedding is combined with sequence clustering with the use of $K$-means algorithm, which prevents from storing whole similarity matrix and keeps memory requirements under control. Both MAFFT and Clustal Omega use tuple matching for similarity calculation.

While MAFFT and Clustal Omega are computationally applicable for families of even 100\,000 proteins, we show that quality of results for such problems is often unsatisfactory. In this paper we present FAMSA, a progressive multiple alignment algorithm especially suitable for large sets of sequences. Pairwise similarities are established, similarly to Kalign-LCS, on the basis of longest common subsequences (LCS). Unlike MAFFT and Clustal Omega, FAMSA calculates all pairwise similarities, which is efficient due to utilisation of multithreaded, bit-parallel LCS algorithm suited for AVX extensions\cite{I2015} of modern processors. Employing memory-saving single-linkage algorithm~\cite{S1973} for guide tree construction, reduces memory requirements of the first stage to $O(k)$. An important factor contributing to the computational scalability of FAMSA is a novel, in-place algorithm of profile alignment which prevents memory reallocations during the progressive stage. As a result, FAMSA is the fastest and most memory-efficient alignment software when large protein families are of interest. The predominance was observed on sets ranging from thousands to a half million of sequences.

The efficiency of FAMSA comes with superior accuracy. This is thanks to a number of algorithm features. They include using LCS for similarity measurement, MIQS substitution matrix~\cite{YT2014}, and a correction of gap penalties inspired by MUSCLE~\cite{E2004}. The penalties are additionally adjusted to the set size which is a novel technique in alignment software, particularly profitable for large sets of sequences. Misalignments during progressive stage are fixed with a use of refinement scheme similar to the one included in QuickProbs~2~\cite{GD2015}. Consequently, when sets of a few thousands or more sequences are of interest, FAMSA is significantly more accurate than any other algorithm. Importantly, the difference increases with growing number of sequences. E.g., for sets exceeding 25\,000 proteins, FAMSA properly aligned 35\% and 25\% more columns than the most accurate variants of MAFFT and Clustal Omega. When largest benchmark family containing 415\,519 sequences was investigated, the advance was even more remarkable---FAMSA successfully restored 4 times more columns than competitors, at a fraction of required time and memory.

Scalability of FAMSA was assessed on extHomFam, a new benchmark generated analogously to HomFam~\cite{SWD2011} by enriching Homstrad~\cite{Mizuguchi1998}  with families from PFam database~\cite{Punta2012}. 
%It contains 380 sequence sets of sizes ranging from 218 to 415,519. 
It contains 380 sets of sizes ranging from 218 to 415\,519 sequences. 
The abundance of numerous protein families ($k>10\,000$) makes extHomFam particularly representative for large-scale alignment problems, which are of crucial importance in the face of recent advances in high throughput sequencing.

\section*{Methods}

%\subsection{General scheme of the algorithm}
FAMSA, similarly to other progressive algorithms, is composed of four stages: 
\begin{enumerate}
\item Calculation of pairwise similarities,
\item Determination of a guide tree,
\item Progressive profile merging according to the guide tree order,
\item Optional iterative refinement of the final profile.
\end{enumerate}
Detailed descriptions of the algorithm stages together with analyses of time and space complexities are given in the following subsections.

\subsection*{Pairwise similarities calculation}
To determine the pairwise similarities of sequences in the input set we use the length of a longest common subsequence (LCS).
The choice was motivated by the promising results of LCS application to this task in the former studies~\cite{DDG2014,PH2012}.
Given two sequences $A$ and $B$, the length of an LCS is the maximal number of perfectly matching columns. This can be considered as an estimation of true pairwise alignment.
To compensate the effect of LCS length being larger for longer sequences, the value is normalised by the indel distance (the number of single-symbol insertions and deletions necessary to transform one sequence to another). 
This distance approximates the misalignment cost, i.e., the number of gaps in the alignment, in which only perfect matches are allowed.
To penalise the differences between two sequences more than reward the similarities, indel distance is squared, as in the former work on Kalign-LCS~\cite{DDG2014}:
$$\mathit{similarity}(A, B) = \frac{\mathit{LCS\_len}(A, B)}{\mathit{indel}(A, B)^2}.$$

The LCS length can be computed using a straightforward dynamic programming (DP) rule.~\cite{G1997}
Thanks to the internal properties of the DP matrix, the calculation can be made using the bit-parallel approach, in which $w$ cells are computed at a time ($w$ is a computer word size equal to 64 in modern architectures)~\cite{H2004}. The indel distance for the sequences $A$ and $B$ can be directly derived from the LCS length according to the formula:
$$\mathit{indel}(A, B) = |A| + |B| - 2\times \mathit{LCS\_len}(A, B),$$
where $|S|$ denotes the length of the $S$ sequence.
The time complexity of the pairwise similarity calculation is:
$$O\left(\frac{|A| |B|}{w}\right),$$
under reasonable assumption that $w$ is comparable or smaller to the longer sequence length.

As modern computers are equipped with multi-core processors, FAMSA distributes the calculation of LCS lengths for different pairs of sequences to several computing threads. Additionally, presented software makes use of vector operations provided by technologies like SSE, AVX, AVX\,2~\cite{I2015} which are supported by contemporary processors. This allows multiple pairs of sequences to be processed simultaneously by the same thread. Assuming $t$ processing threads, $a$ words in a single AVX vector (2 for AVX, 4 for AVX2, and 8 for announced AVX-512), and $n$ being a sequence length, the total time complexity of the first stage can be expressed as:
$$O\left(\frac{n^2}{w} \times \frac{k^2}{ta}\right).$$

The utilisation of massively parallel architectures has become widespread in computationally demanding tasks. As FAMSA was designed for analysing large protein families, it allows massively parallel devices like graphics processors to be employed for calculation of pairwise similarities. The procedure is implemented in OpenCL, therefore it is suitable for GPUs produced by all main vendors including NVidia and AMD. Distributing LCS computation over thousands of graphics processor threads further increases throughput of the first FAMSA stage. Yet, as is shown in the experimental part of the article, even without the aid of OpenCL, FAMSA is able to process hundreds of thousands of proteins in very short time.

\subsection*{Determination of the guide tree}
A number of algorithms for guide tree construction have been developed, e.g., NJ~\cite{SN1987}, UPGMA~\cite{SM1958}, single-linkage~\cite{FLPSZ1951}.
FAMSA employs the latter, which is motivated by following reasons:
\begin{itemize}
\item it can be computed incrementally, i.e., without storing the complete similarity matrix,
\item it is very fast, i.e., can be completed in $O(k^2)$ time using the SLINK algorithm~\cite{S1973},
\item it gave superior results in former studies~\cite{PH2012}.
\end{itemize}
To benefit from the incremental property of SLINK, first two stages of FAMSA are performed simultaneously, which restricts memory footprint. Particularly, tree generation requires only $O(k)$ space in contrast to $O(k^2)$ needed by other guide tree construction algorithms like UPGMA. This is of crucial importance when huge protein families are investigated.

\subsection*{Progressive construction of the alignment}
Progressive construction stage requires $O(k)$ profile alignments, each computed with a use of dynamic programming. At least half of these alignments are degenerated cases in which one or both profiles consist of a single sequence. As dynamic programming implementation can be simplified in those cases, we prepared specialised variants of the general DP procedure. This gave remarkable computation time savings for huge datasets, in which due to the structure of a guide tree, the majority of profile alignments are made against a single sequence.

\begin{figure}
\iffigsinpdf
\centerline\includegraphics{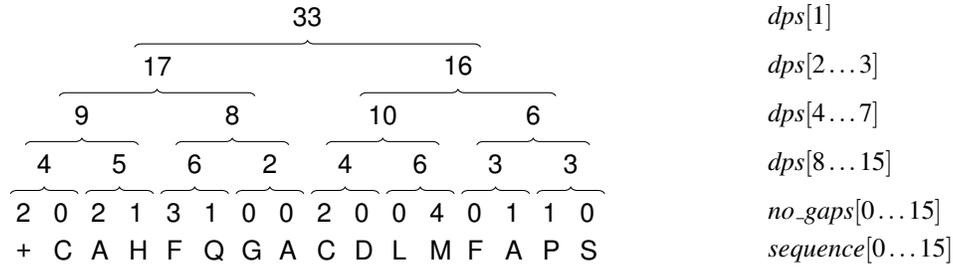}
\else
\centering
\begin{tikzpicture}[>=stealth,x=0.5cm,y=0.5cm]
%\small
\sffamily
\draw(0,0)[anchor=west] node{+};
\draw(1,0)[anchor=west] node{C};
\draw(2,0)[anchor=west] node{A};
\draw(3,0)[anchor=west] node{H};
\draw(4,0)[anchor=west] node{F};
\draw(5,0)[anchor=west] node{Q};
\draw(6,0)[anchor=west] node{G};
\draw(7,0)[anchor=west] node{A};
\draw(8,0)[anchor=west] node{C};
\draw(9,0)[anchor=west] node{D};
\draw(10,0)[anchor=west] node{L};
\draw(11,0)[anchor=west] node{M};
\draw(12,0)[anchor=west] node{F};
\draw(13,0)[anchor=west] node{A};
\draw(14,0)[anchor=west] node{P};
\draw(15,0)[anchor=west] node{S};

\draw(0,1)[anchor=west] node{2};
\draw(1,1)[anchor=west] node{0};
\draw(2,1)[anchor=west] node{2};
\draw(3,1)[anchor=west] node{1};
\draw(4,1)[anchor=west] node{3};
\draw(5,1)[anchor=west] node{1};
\draw(6,1)[anchor=west] node{0};
\draw(7,1)[anchor=west] node{0};
\draw(8,1)[anchor=west] node{2};
\draw(9,1)[anchor=west] node{0};
\draw(10,1)[anchor=west] node{0};
\draw(11,1)[anchor=west] node{4};
\draw(12,1)[anchor=west] node{0};
\draw(13,1)[anchor=west] node{1};
\draw(14,1)[anchor=west] node{1};
\draw(15,1)[anchor=west] node{0};

\draw[snake=brace] (0.1,1.5) -- (1.9,1.5); \draw[anchor=south](1,1.8) node{4};
\draw[snake=brace] (2.1,1.5) -- (3.9,1.5); \draw[anchor=south](3,1.8) node{5};
\draw[snake=brace] (4.1,1.5) -- (5.9,1.5); \draw[anchor=south](5,1.8) node{6};
\draw[snake=brace] (6.1,1.5) -- (7.9,1.5); \draw[anchor=south](7,1.8) node{2};
\draw[snake=brace] (8.1,1.5) -- (9.9,1.5); \draw[anchor=south](9,1.8) node{4};
\draw[snake=brace] (10.1,1.5) -- (11.9,1.5); \draw[anchor=south](11,1.8) node{6};
\draw[snake=brace] (12.1,1.5) -- (13.9,1.5); \draw[anchor=south](13,1.8) node{3};
\draw[snake=brace] (14.1,1.5) -- (15.9,1.5); \draw[anchor=south](15,1.8) node{3};

\draw[snake=brace] (0.5,2.8) -- (3.5,2.8); \draw[anchor=south](2,3.1) node{9};
\draw[snake=brace] (4.5,2.8) -- (7.5,2.8); \draw[anchor=south](6,3.1) node{8};
\draw[snake=brace] (8.5,2.8) -- (11.5,2.8); \draw[anchor=south](10,3.1) node{10};
\draw[snake=brace] (12.5,2.8) -- (15.5,2.8); \draw[anchor=south](14,3.1) node{6};

\draw[snake=brace] (1.4,4.1) -- (6.6,4.1); \draw[anchor=south](4,4.4) node{17};
\draw[snake=brace] (9.4,4.1) -- (14.6,4.1); \draw[anchor=south](12,4.4) node{16};

\draw[snake=brace] (3.4,5.4) -- (12.6,5.4); \draw[anchor=south](8,5.7) node{33};

\draw[anchor=west] (20,6.2) node{$\mathit{dps}[1]$};
\draw[anchor=west] (20,4.9) node{$\mathit{dps}[2\ldots 3]$};
\draw[anchor=west] (20,3.6) node{$\mathit{dps}[4\ldots 7]$};
\draw[anchor=west] (20,2.3) node{$\mathit{dps}[8\ldots 15]$};
\draw[anchor=west] (20,1.0) node{$\mathit{no\_gaps}[0\ldots 15]$};
\draw[anchor=west] (20,0.0) node{$\mathit{sequence}[0\ldots 15]$};

\end{tikzpicture}
\fi
\caption{Illustration of gapped sequence representation of 
%{\sffamily CAHFQGACDLMFAPS} 
{\small\sffamily -- -- C A -- -- H -- F -- -- -- Q -- G A C -- -- D L M -- -- -- -- F A -- P -- S}.
The `+' symbol is a guard present to simplify the implementation.
The values of $\mathit{dps}$ are computed according to the rule: $\mathit{dps}[i] = \mathit{dps}[2i] + \mathit{dps}[2i+1]$, if the necessary cells are present.
Otherwise they are calculated on the $\mathit{no\_gaps}$ and $\mathit{sequence}$ vectors.
E.g., $\mathit{dps}[8]$ is the number of symbols in $\mathit{sequence}[0\ldots 1]$ (equal 2) incremented by the number of gaps present just before these symbols, i.e., $\mathit{no\_gaps}[0]$ and $\mathit{no\_gaps}[1]$.
}
\label{fig:gapped-sequence}
\end{figure}

Several improvements to the classical computation rule were introduced in FAMSA to increase alignment quality as well as the processing speed.
They were possible thanks to the internal profile representation composed of three arrays storing:
\begin{itemize}
\item occurrence counters of each alphabet symbol in consecutive columns (occupying $32 n^*$ computer words, with $n^*$ being the profile length),
\item costs of alignment of consecutive columns to each possible alphabet symbol (occupying also $32 n^*$ computer words), 
\item sequences in the \emph{gapped representation}.
\end{itemize}
While two former components were previously employed by alignment algorithms, e.g., Kalign, the gapped representation is, to the best of our knowledge, a novel technique.
In this representation, for each sequence, two equal-sized arrays are stored: (\emph{i}) sequence symbols, (\emph{ii}) a number of gaps present before the corresponding sequence symbol.
Moreover, to quickly localise a symbol in a column, as well as to insert or remove gaps, a dynamic position statistics are stored in an additional array.
The space for the gapped sequence is approximately $13$ times the length of the sequence (see Figure~\ref{fig:gapped-sequence} for example).
The proposed profile representation allows a dynamic programming matrix to be computed rapidly and is memory frugal.
The DP computation step for a pair of profiles takes
$$O(n_1 n_2 \sigma)$$
time, where $n_1$, $n_2$ are the input profile lengths and 
$\sigma$ is the alphabet size (equals 32 to represent twenty amino acids and several special symbols).

The presence of gapped representation is especially beneficial when families containing tens of thousands proteins are investigated. Other aligners construct new profile by copying sequences symbol by symbol with occasional gap insertions, which starts to be a bottleneck for large-scale analyses. This is not the case in FAMSA in which whole sequences are moved from the input profiles to the new one and gaps are rapidly inserted by updating gap counters in corresponding arrays.
The time of construction of a new profile is:
$$O(k_\text{o} + n_\text{o}\sigma + k_1 (n_\text{o} - n_1) \log n + k_2 (n_\text{o} - n_2) \log n),$$
where $k_1$ and $k_2$ are number of sequences in both profiles, $k_\text{o} = k_1 + k_2$, and $n_\text{o}$ is the resulting profile length.
The overall time of all profile constructions is thus:
$$O(k^2 + n_\text{f}k\sigma + (n_\text{f}-n)k\log n) = O(k^2 + n_\text{f}k(\sigma + \log n)),$$
where $n_\text{f}$ is the final profile length.

Adding the time of DP matrix calculation gives the total time of this stage:
$$O(k n_\text{f}^2\sigma + k^2 + n_\text{f}k(\sigma+\log n)) = O(k^2 + n_\text{f}^2k\sigma).$$

As profile alignments in the bottom part of the guide tree are independent, they can be performed in parallel.
Therefore, to improve the computation time, FAMSA distributes profile alignments over multiple threads.
It would also be possible to parallelise the dynamic programming computation and construction of a single profile. We expect it to be particularly beneficial for families of million and more proteins. Nevertheless, we refrained from this in the current FAMSA version due to implementation complications and lack of that large sets in existing databases.

\subsection*{Gap types and costs determination}
Among numerous amino acid substitution matrices for dynamic programming calculation, we selected MIQS due to superior results reported in the recent study~\cite{YT2014}.
The gap costs are determined according to the classic affine penalty function, with a distinction between terminal and non-terminal gap open and gap extension costs, similarly to Kalign~\cite{LS2005,LFS2009} or MUSCLE~\cite{E2004}.
Particularly, four types of gaps are used:
\begin{itemize}
\item gap\_terminal\_open ($T_\text{o}$)---opens a sequence at the left end or opens a contiguous series of gaps at the right end of a sequence,
\item gap\_terminal\_extension ($T_\text{e}$)---extends a series of gaps inserted at the beginning or end of a sequence,
\item gap\_open  ($G_\text{o}$)---opens a contiguous series of gaps inserted within a sequence,
\item gap\_extension ($G_\text{e}$)---extends a contiguous series of gaps inserted within a sequence.
\end{itemize}

While determination of the number of gaps and their types is straightforward in pairwise alignment, it becomes problematic in MSA. 
%Due to the fact that before aligning two profiles $X$ and $Y$, their sequences may already contain gaps, the insertion of a single column of gaps (or the first column of gaps in a contiguous series of gaps) is not always equivalent to the insertion of gap\_opens exclusively.
Due to the fact that before aligning two profiles their sequences may have already contained gaps, the insertion of a column of gaps (either a single one or as the first one in a contiguous series of columns with gaps) is not always equivalent to the insertion of gap\_opens exclusively.
%Inserting only gap\_opens would result in the cost overestimation.
Inserting only gap\_opens would result in an overestimation of their number.
%That is why some corrections of the inserted gaps should be incorporated to lower the total gap penalty. 
%That is why we correct the inserted gaps ....should be incorporated to lower the total gap penalty.
%The estimated cost of the alignment should take into account some modifications of the inserted gaps.
That is why types of gaps within a column should be corrected.
%We consider the following situations in Figure~\ref{fig:gap-cor}.

\begin{figure}
\iffigsinpdf
\centerline\includegraphics{famsa-figure1}
\else
\newcommand{\Gto}[1][blue]{\color{#1}$\mathsf{T_o}$}
\newcommand{\Gte}[1][blue]{\color{#1}$\mathsf{T_e}$}
\newcommand{\Go}[1][blue]{\color{#1}$\mathsf{G_o}$}
\newcommand{\Ge}[1][blue]{\color{#1}$\mathsf{G_e}$}
\newcommand{\Gap}[1][red]{\color{#1}--}

\newcommand{\Seq}[8]{
\setlength{\tabcolsep}{0em}
\begin{tabular}{p{1cm}p{0.6cm}p{0.6cm}p{0.6cm}p{0.6cm}p{0.6cm}p{0.6cm}p{0.6cm}}
#1 & \centerline{#2} & \centerline{#3} & \centerline{#4} & \centerline{#5} & \centerline{#6} & \centerline{#7} & \centerline{#8}
\end{tabular}
}
\centering
\begin{tikzpicture}[>=stealth,x=0.6cm,y=0.5cm]
%\small
\sffamily
% Left part
\draw(2, 0)[black,anchor=west] node{\Seq{$S_1$:}{C}{D}{E}{\Gap}{\Gto}{\Gte}{\Gte}};
\draw(2,-1)[black,anchor=west] node{\Seq{$S_2$:}{\Gto}{\Gte}{\Gte}{\Gap}{\Gte}{H}{I}};
\draw(2,-2)[black,anchor=west] node{\Seq{$S_3$:}{C}{D}{\Go}{\Gap}{\Ge}{H}{I}};
\draw(2,-3)[black,anchor=west] node{\Seq{$S_4$:}{C}{D}{E}{\Gap}{\Go}{\Ge}{I}};
\draw(2,-4)[black,anchor=west] node{\Seq{$S_5$:}{C}{D}{E}{\Gap}{F}{H}{I}};
\draw(2,-5)[black,anchor=west] node{\Seq{$S_6$:}{C}{D}{\Gto}{\Gap}{\Gte}{\Gte}{\Gte}};
\draw[thick] (1.5,-5.4)--(12.0,-5.4);
\draw(2,-6.5)[black,anchor=west] node{\Seq{$S_7$:}{C}{\Go}{\Ge}{Q}{F}{H}{I}};

\draw(2,0.8)--(1.8,0.7)--(1.8,-4.9)--(2,-5);
\draw(2,-5.7)--(1.8,-5.8)--(1.8,-6.4)--(2,-6.5);
\draw(0.5,-2.0)[black,anchor=west] node{$X$};
\draw(0.5,-6.0)[black,anchor=west] node{$Y$};

% Right part
\draw(18, 0)[black,anchor=west] node{\Seq{$S_1$:}{C}{D}{E}{\Gto[red]}{\Gte[red]}{\Gte}{\Gte}};
\draw(18,-1)[black,anchor=west] node{\Seq{$S_2$:}{\Gto}{\Gte}{\Gte}{\Gte[red]}{\Gte}{H}{I}};
\draw(18,-2)[black,anchor=west] node{\Seq{$S_3$:}{C}{D}{\Go}{\Ge[red]}{\Ge}{H}{I}};
\draw(18,-3)[black,anchor=west] node{\Seq{$S_4$:}{C}{D}{E}{\Go[red]}{\Ge[red]}{\Ge}{I}};
\draw(18,-4)[black,anchor=west] node{\Seq{$S_5$:}{C}{D}{E}{\Go[red]}{F}{H}{I}};
\draw(18,-5)[black,anchor=west] node{\Seq{$S_6$:}{C}{D}{\Gto}{\Gte[red]}{\Gte}{\Gte}{\Gte}};
\draw[thick] (17.5,-5.4)--(28.0,-5.4);
\draw(18,-6.5)[black,anchor=west] node{\Seq{$S_7$:}{C}{\Go}{\Ge}{Q}{F}{H}{I}};

\draw(18,0.8)--(17.8,0.7)--(17.8,-4.9)--(18,-5);
\draw(18,-5.7)--(17.8,-5.8)--(17.8,-6.4)--(18,-6.5);
\draw(16.5,-2.0)[black,anchor=west] node{$X$};
\draw(16.5,-6.0)[black,anchor=west] node{$Y$};
\end{tikzpicture}
%\end{center}
\fi
\caption{Example of how gap columns are inserted during profile alignment}
\label{fig:gap-cor}
\end{figure}

%\begin{figure*}
%\centerline{\includegraphics[width=\textwidth]{gap_cor}}
%\caption{Example of how gap columns are inserted during profile alignment}
%\label{fig:gap-cor}
%\end{figure*}

%An alignment of two profiles $X$ and $Y$.
%The situations that we consider are shown in Figure~\ref{fig:gap-cor}, where an alignment of two profiles $X$ and $Y$ is investigated.
%A column of gaps is to be inserted into profile $X$ (left part of the figure). 
%The proper types of gaps together with corrected gaps at the neighboring column are shown in the right part of the figure.

In Figure~\ref{fig:gap-cor} an exemplary alignment of two profiles $X$ and $Y$ is shown.
A column of gaps is to be inserted into profile $X$ (left part of the figure). 
The proper types of gaps together with corrected gaps at the neighboring column are shown in the right part of the figure.
While correcting gaps we consider  the following situations:
\begin{itemize}
\item $S_1$: there is a gap\_terminal\_open at the right side of the inserted one; hence, the inserted gap should be gap\_terminal\_open, and the following gap should be transformed into gap\_terminal\_extension,
\item $S_2$: there is a gap\_terminal\_extension at the left side of the inserted one; hence, the inserted gap should also be gap\_terminal\_extension,
\item $S_3$: the inserted gap is to be placed into the gap series, so it should be gap\_extension,
\item $S_4$: there is a gap\_open at the right side of the inserted one, hence, the inserted gap should be gap\_open, and to prevent the occurrence of two gap\_opens one after the other, the second gap should turn into a gap\_extension,
\item $S_5$: the inserted gap is to be placed within the series of residues as the only gap, so it should be gap\_open,
\item $S_6$: there is a gap\_terminal\_open at the left side of the inserted one, hence, the inserted gap should be gap\_terminal\_exten\-sion.
\end{itemize}

Optimising gap parameters and recognising its influence on alignment accuracy is still the subject of intensive studies~\cite{E2009}.
Various techniques have been proposed, e.g., adding a bonus score to a gap cost to force to align distantly related sequences~\cite{LFS2009}. 
%Therefore, to counteract this tendency, all gap costs (i.e., gap opens and gap extensions, both terminal and nonterminal) are multiplied by a factor related to the number of sequences in the input collection.
In our research all gap costs (i.e., gap opens and gap extensions, both terminal and non-terminal) are multiplied by a factor related to the number of sequences in the input collection.
This prevents unnecessary widening of alignments of large collections.
The scaling factor is calculated as:
$$g_\text{scale} = 1 + \frac{\log(k/ g_\text{l})}{g_\text{d}},$$
where $g_\text{l}$ and $g_\text{d}$ are two constants set by default to 45 and 7 (values chosen experimentally).

%The application of gap scaling leads to another optimization of the dynamic programming calculation rule.  
The application of gap corrections and scaling leads to another modification of the traditional approach. 
%To speedup the computations, when the current DP matrix cell is being determined, it is usually assumed that the insertion of a gap column to the first profile cannot be immediately followed by the insertion of a gap column to the second profile.
It is usually assumed that the insertion of a gap column to the first profile cannot be immediately followed by the insertion of a gap column to the second profile.
Under some assumptions about the gap costs and substitution matrix values, it can be proved to be reasonable, i.e., such situation never leads to the optimal alignment.
Nevertheless, this is not true if the gap correction is applied.
%Therefore, it is checked whether consecutive insertions of gap columns to both profiles renders higher score in the DP matrix.
%\footnote{
%AG: Ogolnie nie bylo dla mnie jasne, czy omawiane nastepstwo we wstawianiu gapow dotyczy czasu czy przestrzeni. Nie jestem tez pewny, czy to optymalizacja czasowa, czy jakosciowa (a moze to tylko zniwelowanie bledu wprowadzanego przez korekcje gapow?). Generalnie troche byl ten fragment niezrozumialy dla niezorientowanego czytelnika. Pozmienialem go wedlug swojego wyczucia, ale moglem cos zrozumiec na opak. Stary tekst zostawilem zakomentowany jakby co.
%
%SD: Mniej wiecej jest ok, ale to ``higher score'' kurcze mnie martwi. 
%Generalnie ciezko sie czyta te opisy, bo nie mowimy co to sa za wartosci w macierzy DP i tylko czasami rzucamy takimi haslami. Czytelnika to na pewno oglupi.
%}
Therefore, it is checked whether consecutive insertions of gap columns to both profiles render a higher-scored alignment\footnote{The profile alignment score is the summed alignment score of each sequence pair.}.

%One more optimization is extending the DP calculation rule.
%To speedup the computation when the value of the current cell of the DP matrix is calculated, it is usually assumed that just after inserting a gap column into one profile, a gap column cannot be inserted into an other profile.
%Under some assumptions about the gap costs and substitution matrix values, it can be proved that this is reasonable, i.e., such situation never leads to the optimal alignment.
%Nevertheless, this is not true if the gap correction is applied.
%Therefore, in FAMSA it is checked whether such insertion of gaps in one profile just after insertion of gaps in another one gives better score in the DP matrix.

\subsection*{Iterative refinement}
The idea of an iterative refinement is to correct misalignments made in the early phase of the profile alignment.
Several algorithms were proposed for this task, like REFINER~\cite{CLPPTB2006} or the methods implemented in MMSA~\cite{PH2012}, MSAProbs~\cite{LSM2010}.
In our recent paper~\cite{GD2015} we investigated this problem showing that for sufficiently large collections of sequences the classical methods did not work.
We also proposed a column-oriented refinement to improve the quality of alignments for collections up to 1000 sequences.
In this approach, the algorithm scans the profile to localise columns that contain at least one gap.
Then, it randomly selects one of such columns and splits the profile into two subprofiles, depending on the gap presence in the selected column. Empty columns are removed afterwards and subprofiles are realigned.
%
%Unfortunately, our gap correction does not work well in the refinement.
%Namely, when we remove some gap-only columns some gaps are misclassified as gap extensions when they should be gap opens.
%To solve this problem, a huge computational effort would be necessary so we resigned form this.
%Finally, if new alignment has higher SP score\footnote{Znowu tu jest SP score i to bez definicji boli} than the original one, it is accepted as the current solution.
% (similarly like in MUSCLE~\cite{E2004}).
Finally, if new alignment is scored higher than the original one, it is accepted as the current solution.

To simplify the time complexity analysis of refinement, we assume the input and the output profiles to be of comparable lengths (which is usually the case). A single refinement iterations requires then
$$O(kn_\text{f} + n_\text{f}^2\sigma + k(n_\text{f} - n_\ast)\log n)$$
time, with $n_\ast$ being the length of the shorter of the two profiles obtained after splitting the original profile.

Preliminary analyses showed refinement to be particularly beneficial for smaller sets of sequences. Due to this reason and to improve the processing time of large protein families, the refinement is applied only for $k \le 1000$. Number of iterations was experimentally set to 100.

\section*{Results}

\subsection*{Benchmark selection}
An assessment of MSA algorithms was performed using benchmark datasets. The presence of high-quality, manually curated reference alignments allowed supervised accuracy measures to be calculated. Those were sum-of-pairs (SP) and total-column (TC) scores defined as fractions of correctly aligned symbol pairs and columns, respectively.    

Our aim was to propose an efficient and robust algorithm for the alignment of thousands of proteins.
The largest available benchmarks contain sets of at most hundreds of sequences, with an exception of HomFam introduced by Sievers \emph{et al.}\cite{SWD2011}.
HomFam consists of 92 families constructed by extending Homstrad reference alignments (only those having 5 or more sequences were taken into account) with corresponding families from PFam database. This protocol results in large benchmark sets: 18 of them consists of more than 10\,000 members (the number of reference sequences ranges from a few to a few tens, though).
 
Since 2011, when HomFam was introduced, PFam and Homstrad databases have grown significantly. Therefore, to carry out more extensive experiments, we present a new benchmark named extHomFam. It was constructed according to the HomFam generation protocol with several modifications. 399 high-quality alignments containing at least 3 proteins were selected from Homstrad (ver.\ 1 Apr 2015). By decreasing the threshold from 5, we aimed at obtaining a larger benchmark than original HomFam. Taking into account also two-protein families would increase extHomFam to 1\,013 sets, at the cost of positively biasing TC score (with two reference sequences it becomes equal to SP), therefore pairwise-only alignments were excluded from the consideration. After that, selected Homstrad sets were enriched with corresponding PFam (ver.\ 28) families.
%(unlike HomFam, one-to-one mapping was not required). 
After removing duplicated sequences, sets of less than 200 proteins were filtered out giving final benchmark of 380 families. For convenience extHomFam was divided at thresholds $k=$ 4\,000; 10\,000, and 25\,000 to obtain subsets named \emph{small}, \emph{medium}, \emph{large}, and \emph{extra-large}. Note, that sets of $\sim$1000 sequences are usually referred to in the literature as large---our naming convention is to show size diversity. \emph{ABC\_tran}, the most numerous set in extHomfam contains 415\,519 sequences, which is the largest benchmark protein family available. 
  
The scalability of algorithms was evaluated on 53 largest extHomFam families containing at least 30\,000 sequences each. %(the largest $k$ on performance charts). 
These sets were recursively downsampled to desired sizes with a guarantee of preserving sequences from reference alignments. This scheme has a valuable property of smaller sets being contained in larger ones which reduces results variability. 

To evaluate the performance of the presented algorithm on smaller alignment problems, classic benchmarks, i.e., BAliBASE~\cite{Thompson1999}, PREFAB~\cite{E2004}, OXBench-X~\cite{Raghava2003}, and SABmark~\cite{Walle2005}, were also considered in the experiments.

\subsection*{Competitive algorithms and system setup}
From among numerous sequence alignment algorithms, only those able to handle families of thousands of sequences were investigated on HomFam and extHomFam. Those were MUSCLE~\cite{E2004}, Kalign~2\cite{LFS2009}, Kalign-LCS~\cite{DDG2014}, Clustal Omega~\cite{SWD2011}, and MAFFT~\cite{Katoh2013}. The latter was analysed in default configuration in which it calculates $O(k^2)$ pairwise similarities, as well as -parttree and \mbox{-dpparttree} modes especially suited for large sets of sequences due to lower computational requirements. Clustal Omega was executed with default parameters and with two combined iterations (-iter2) which was shown to give superior results in the previous studies~\cite{SWD2011,SDWH2013}.
% ($O(k \log k)$).  - nie jest jasne czego to zlozonosc, wiec moze olac te informacje
MUSCLE in default mode was unfeasible for immense protein families, hence -maxiters2 variant was also considered.
Details on execution parameters and program versions are given in the Supplementary material.

The experiments on smaller benchmarks (BAliBASE, PREFAB, OXBench-X, SABmark) concerned also top consistency-based algorithms: MSAProbs~\cite{LSM2010}, QuickProbs~\cite{GD2014,GD2015}, and GLProbs~\cite{YCYZLT2015}.

For the experiments we used the workstation equipped with two 12-core Intel Xeon E5-2670v3 processors (clocked at 2.3\,GHz), Nvidia Quadro M6000 graphic card (3072 cores clocked at 1.0\,GHz), and 128\,GB RAM. To investigate the behavior of the algorithms on modern workstations and servers containing from a few to several tens of cores, all methods were run with 8 computing threads, unless stated otherwise. FAMSA was run in the CPU mode, except for the experiment on the algorithm scalability w.r.t. the number of CPU cores, where GPU variant was additionally investigated.

\subsection*{HomFam and extHomFam benchmark evaluation}

Following~Sievers \emph{et al.}\cite{SWD2011}, HomFam was divided into three parts depending on the family size. As Table~\ref{tab:res:HomFam} shows, for $k \leq 3\,000$, FAMSA was inferior to both Clustal Omega configurations and comparable to MAFFT-default. For $3\,000 < k \leq 10\,000$, FAMSA became the first and the second in terms of SP and TC score, respectively. When the last subgroup was investigated, presented algorithm took the lead on both measures revealing its potential for large protein families. Importantly, FAMSA was from several to hundreds times faster than competitors. E.g., it processed entire HomFam in 12 minutes while Clustal-default and MAFFT-default required, respectively, 8h40m and 2h30m. Even larger difference was observed for Clustal-iter2 which completed analyses in 51 hours. MAFFT-parttree and -dpparttree were also inferior to FAMSA, which is especially noteworthy as they calculate only selected pairwise similarities (usually $O(k\log k)$) instead of full matrix ($O(k^2)$).

\begin{table*}[t]
\caption{Comparison of algorithms on HomFam dataset.}
\small
\label{tab:res:HomFam}
\setlength{\tabcolsep}{0.0em}
\begin{tabular*}{\hsize}{@{\extracolsep{\fill}}lcccrcccrcccrcccr}\toprule
Algorithm	&&	
%	\multicolumn{3}{c}{Small}&&	
%	\multicolumn{3}{c}{Medium}&&
%	\multicolumn{3}{c}{Large}&&
%	\multicolumn{3}{c}{All}\\ 
%	&&	
	\multicolumn{3}{c}{$93 \le k \le 3000$}&&	
	\multicolumn{3}{c}{$3000 < k \le 10000$}&&
	\multicolumn{3}{c}{$10000 < k \le 50157$}&&
	\multicolumn{3}{c}{}\\ 
				&&	
	\multicolumn{3}{c}{41 families}&&	
	\multicolumn{3}{c}{33 families}&&	
	\multicolumn{3}{c}{18 families}&&
	\multicolumn{3}{c}{92 families}\\
	\cline{3-5}	\cline{7-9}	\cline{11-13}	\cline{15-17}
			&& SP	& TC	& time	&& SP	& TC	& time	&& SP	& TC	& time	&& SP	& TC	& time	\\
\midrule
FAMSA					&&
	82.6	& 63.5	&\bf30		&&\bf 86.9	& 68.4	&	\bf2:34	&&\bf 73.1	&\bf 50.4	&\bf 8:55	&&
	82.3	& 62.7	&\bf 11:59
	\\
Clustal-iter2		&&
	\bf86.3	&\bf 71.5	& 2:14:18	&& 85.0	&\bf 68.9	&	14:33:00	&& 69.5	& 48.3	& 	34:50:54	&&
	\bf82.5	& \bf 66.0	& 51:38:12	\\
Clustal-default		&&
	85.7	& 70.8	& 27:02	&& 82.7	& 63.9	&	2:24:36	&& 67.6	& 46.4	& 	5:51:10	&&
	81.1	&  63.6	& 8:42:50	\\
MAFFT-default		&&
	81.9	& 64.0	&	2:15	&&	80.8	& 57.6	& 23:55	&&	69.1	& 46.2	& 2:05:52		&&
	79.0	& 58.2		& 2:32:02 \\
MAFFT-parttree	&&
	77.0	& 55.2	& 2:43	&&	72.4	& 46.6	& 15:38	&&	58.0	& 33.0	&	43:08		&&
	71.6	&	47.8	&	1:01:29	\\
MAFFT-dpparttree	&&
	80.3	& 61.2	& 10:40	&&	79.0	& 54.5	& 57:41	&&	63.5	& 37.8	& 1:57:02	&&
	76.5	&	54.2	& 3:05:23	\\
Kalign-LCS			&&
	79.8	& 61.3	& 4:17	&& 80.6	& 57.6	& 1:31:45&&	67.9	& 44.4	& 64:23:14	&&
	77.8	& 56.7	& 65:59:16 \\
Kalign2				&&
	77.4	& 56.2	&	7:04	&&	77.6	& 57.1	& 2:30:45&& 64.8	& 41.6	&	97:48:46	&&
	75.0	& 53.7	&	100:26:35\\
MUSCLE-default		&&	
	72.0	& 53.2	& 35:35:44	&&	---	& ---		&	---	&& ---	& ---		& ---		&&
	---	& ---		& ---	\\
MUSCLE-maxiters2	&&
	71.8	& 51.4	& 12:35		&& 67.1	& 41.6	&	2:27:51	&& 40.6		& 21.6	& 30:35:03			&&
	68.8	& 42.1		& 	33:15:29 \\
\bottomrule
\end{tabular*}
\end{table*}

The experiments on extHomFam confirmed superior accuracy and execution time of FAMSA to scale well with the number of sequences (Figure~\ref{Figure:barplot}; more detailed results are given in Supplementary material). FAMSA was inferior to Clustal-iter2 by a small margin only on \emph{small} subset. For $4\,000 < k \leq 25\,000$ it became the best aligner and, depending on the subset and quality measure, was followed by Clustal, MAFFT, Kalign2, or Kalign-LCS.
% (all these algorithms performed similarly). 
MUSCLE, as well as fast MAFFT variants rendered inferior results (MAFFT-parttree was excluded from Figure~\ref{Figure:barplot} due to significantly worse accuracy than that of -dpparttree). On \emph{extra-large} FAMSA held the lead while MAFFT-dpparttree became the second best algorithm. Kalign2, Kalign-LCS, and MUSCLE did not complete the analyses due to excessive memory or time requirements. Clustal Omega and MAFFT-default failed to process, respectively, one and four largest extHomFam families (missing MAFFT results were taken from -dpparttree variant, though). Advances in SP and TC measures of FAMSA over competing software on \emph{medium}, \emph{large}, and \emph{extra-large} subsets were assessed statistically with a use of Wilcoxon signed-rank test with Bonferroni-Holm correction for multiple testing. The differences are significant at $\alpha=0.05$, $p$-values for all pairwise comparisons can be found in Table~\ref{tab:res:significance}.

\begin{figure*}[!htp]
\centering	
\includegraphics[width=\textwidth]{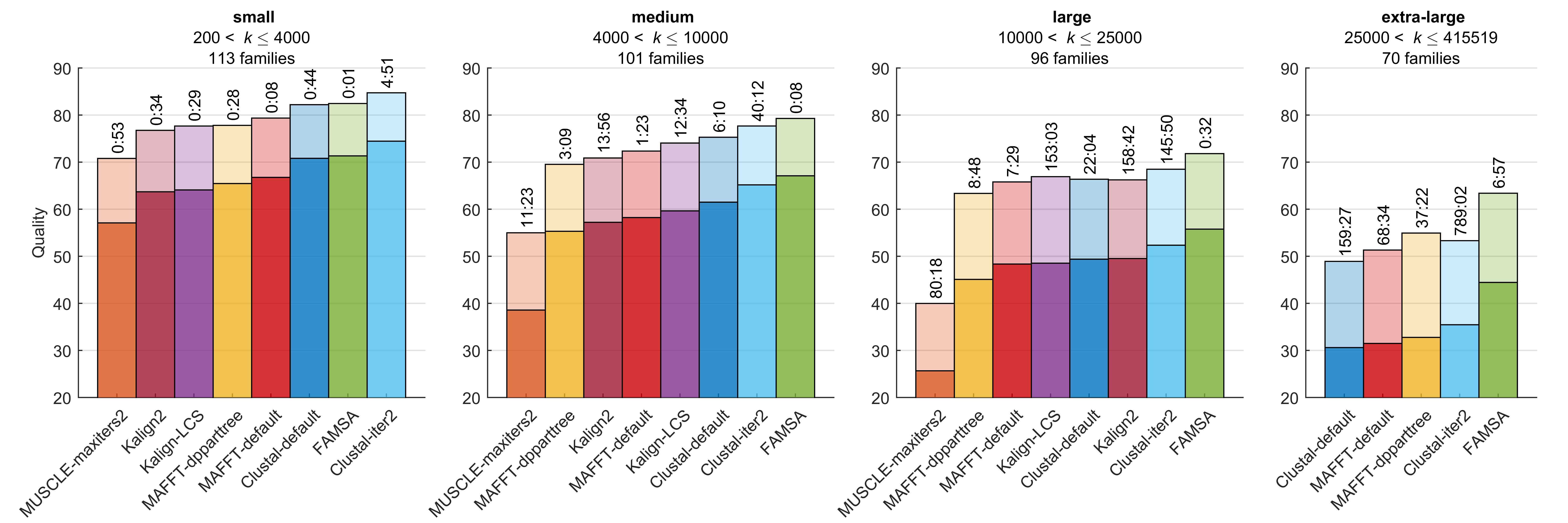}	
\caption{Comparison of alignment software on extHomFam. 
The solid bars (lower) represent TC scores, while the transparent ones (higher)---SP scores.
For each subset algorithms were sorted increasingly according to TC measure. 
Execution times are given above bars in \emph{hours:minutes} format.}
\label{Figure:barplot}
\end{figure*}

\begin{table*}[h]
\caption{Statistical significance of FAMSA advances over selected competitors measured using Wilcoxon signed-rank test with Bonferroni-Holm correction for multiple testing on extHomFam subsets.}
\small
\label{tab:res:significance}
\begin{tabular*}{\hsize}{@{\extracolsep{\fill}}lcccrcccrcccr}\toprule
Algorithm	&&	\multicolumn{2}{c}{medium} && \multicolumn{2}{c}{large}&& \multicolumn{2}{c}{extra-large}\\ 

			\cline{3-4}						\cline{6-7}	\cline{9-10}
					&& SP & TC 						&& SP & TC 					&& SP & TC\\	
\midrule
Clustal-default		&& 0.00003 & 0.00012 		&& 0.00011 & 0.00081		&& $<10^{-5}$ & $<10^{-5}$\\	
Clustal-iter2		&& 0.00971 & 0.00940		&& 0.00502 & 0.01383		&& 0.00065 & 0.00137\\
MAFFT-default		&& $<10^{-5}$ & $<10^{-5}$		&& $<10^{-5}$ & 0.00004		&& $<10^{-5}$ & $<10^{-5}$\\	
MAFFT-dpparttree	&& $<10^{-10}$ & $<10^{-10}$ && $<10^{-7}$ & $<10^{-7}$  && $<10^{-6}$ &  $<10^{-8}$\\
Kalign-LCS			&& $<10^{-6}$ & $<10^{-7}$	&& $<10^{-5}$ &  $<10^{-5}$ 	&& --- & ---\\
\bottomrule
\end{tabular*}
\end{table*}

\begin{figure*}[h]
\centering	
\includegraphics[width=\textwidth]{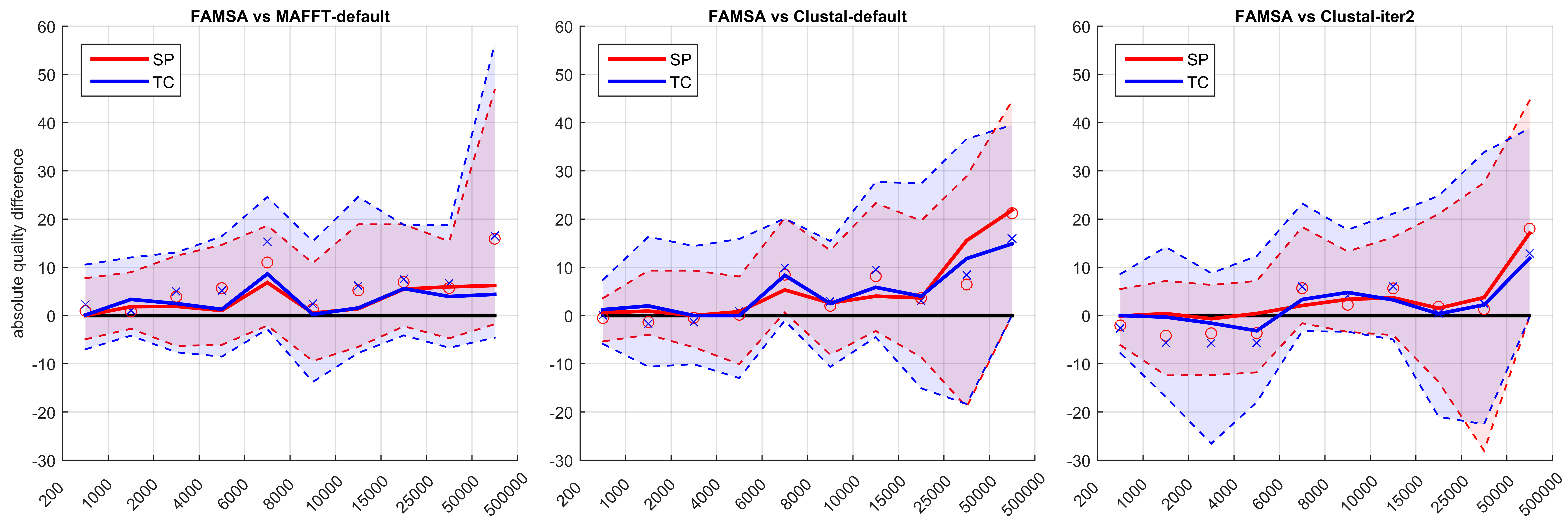}	
\caption{Absolute differences in SP (red) and TC (blue) scores between FAMSA and competing software for extHomFam subsets. Each interval at the horizontal axis contains approximately 35 families. Solid lines represent medians, dashed lines indicate 15th and 85th percentiles (thus, filled areas contain 70\% of observations). Means are additionally given by circular (SP) and cross (TC) markers.}
\label{Figure:percentiles}
\end{figure*}

\begin{figure*}[!t]
	\centering	
	\includegraphics[width=\textwidth]{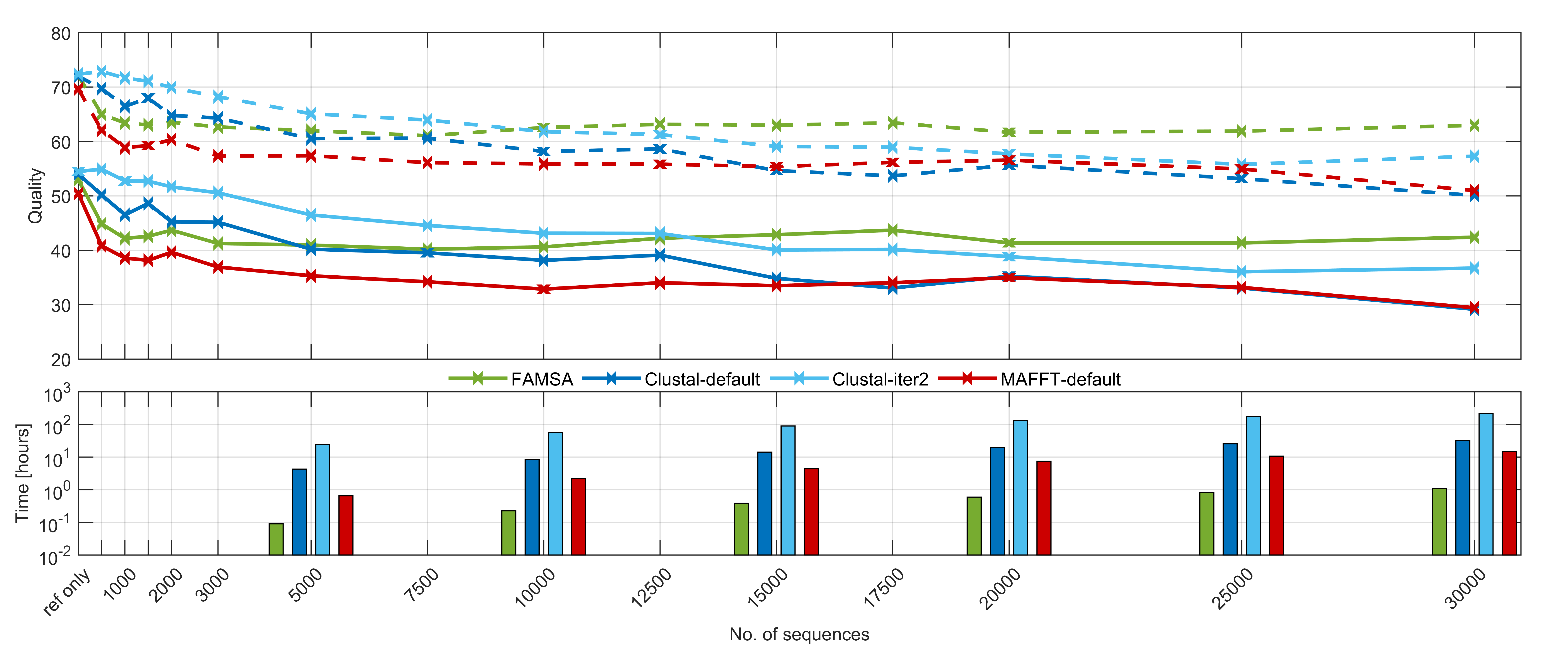}	
	\caption{Scalability of SP (dashed lines) and TC (solid lines) scores with respect to the number of sequences. Experiments were performed on 53 largest extHomFam families randomly resampled to obtain desired set size. Processing times for selected values of $k$ are given as bar plots.}
	\label{Figure:sampling}
\end{figure*}

As Figure~\ref{Figure:barplot} shows, the quality advance of presented software over other algorithms increases for consecutive subsets. For instance, on \emph{extra-large}, FAMSA properly aligned 35\% and 25\% more columns than the most accurate variants of MAFFT and Clustal Omega. More detailed analysis of FAMSA accuracy compared to competitors is given in Figure~\ref{Figure:percentiles}. Four extHomFam categories were further divided into 11 subsets having approximately 35 families. For each interval at $k$ axis, we plotted selected statistical indicators (median, mean, 15th and 85th percentile) of absolute differences in SP and TC measures between FAMSA and other algorithms. Clearly, the number of test cases for which presented software is superior to the competitors, as well as the absolute dominance in quality, increases with growing set size. This observation is supported by the scalability analysis performed on 53 largest families ($k \geq 30\,000$) randomly resampled to obtain less numerous sets. Figure~\ref{Figure:sampling} shows FAMSA to outrun Clustal-default in SP and TC scores when number of sequences exceeds 7500. Clustal-iter2 graphs are crossed for larger sets of sequences, i.e., $k \geq 12\,500$.      

The abundance of extremely large protein families makes extHomFam the most demanding benchmark in terms of computational resources. Beside FAMSA, only MAFFT-dpparttree and -parttree were able to process all its sequences sets. Other algorithms either crashed due to memory requirements or were terminated by us when processing time of a family exceeded 24~hours\footnote{An exception was made only for Clustal-iter2 due to its superior quality results.}. While Clustal-default, MAFFT-default, and MAFFT-dpparttree required, respectively, 188, 78, and 50 hours, FAMSA finished computations in approximately 7 hours and 40 minutes, which corresponds to 25-fold, 10-fold, and 6-fold advance. The extreme case was Clustal-iter2 which needed almost 1000 hours showing its combined iterations to be inapplicable for very large protein families. More detailed analysis of computational scalability of presented algorithm is given in Figure~\ref{Figure:sampling}. It confirms FAMSA to be faster than MAFFT-default and Clustal Omega by approximately order and two orders of magnitude (depending on the parameters of the latter). The efficiency of presented algorithm is thanks to the fast bit-parallel similarity computation and the in-place profile joining. Yet, as FAMSA calculates more distances than Clustal Omega and MAFFT-dpparttree ($O(k^2)$ instead of $O(k\log k)$), one can expect it to exceed competitor execution times for sufficiently large~$k$. To verify this we compared algorithms on \textit{ABC\_tran}, the largest family in extHomFam with 415\,519 proteins. FAMSA processed this set in less then 2 hours. Clustal Omega crashed due to excessive memory requirements after 55 hours of calculations strongly suggesting that the algorithm is dominated by stages other than similarity computation. The different situation was in the case of MAFFT-dpparttree which execution time scaled better with the number of sequences, though was still inferior to FAMSA by a factor of 2.5. Importantly, FAMSA required below 8\,GB of RAM, while MAFFT-dpparttree allocated 47\,GB. To compare, MAFFT-default and Clustal Omega failed to run on 128\,GB machine (the former demanded 318\,GB just for storing the similarity matrix). Concluding, the calculation of all pairwise similarities performed by FAMSA did not prevent it from being the fastest and most memory efficient aligner in the comparison even for immense protein families.

\begin{figure*}[!t]
	\centering	
	\includegraphics[width=\textwidth]{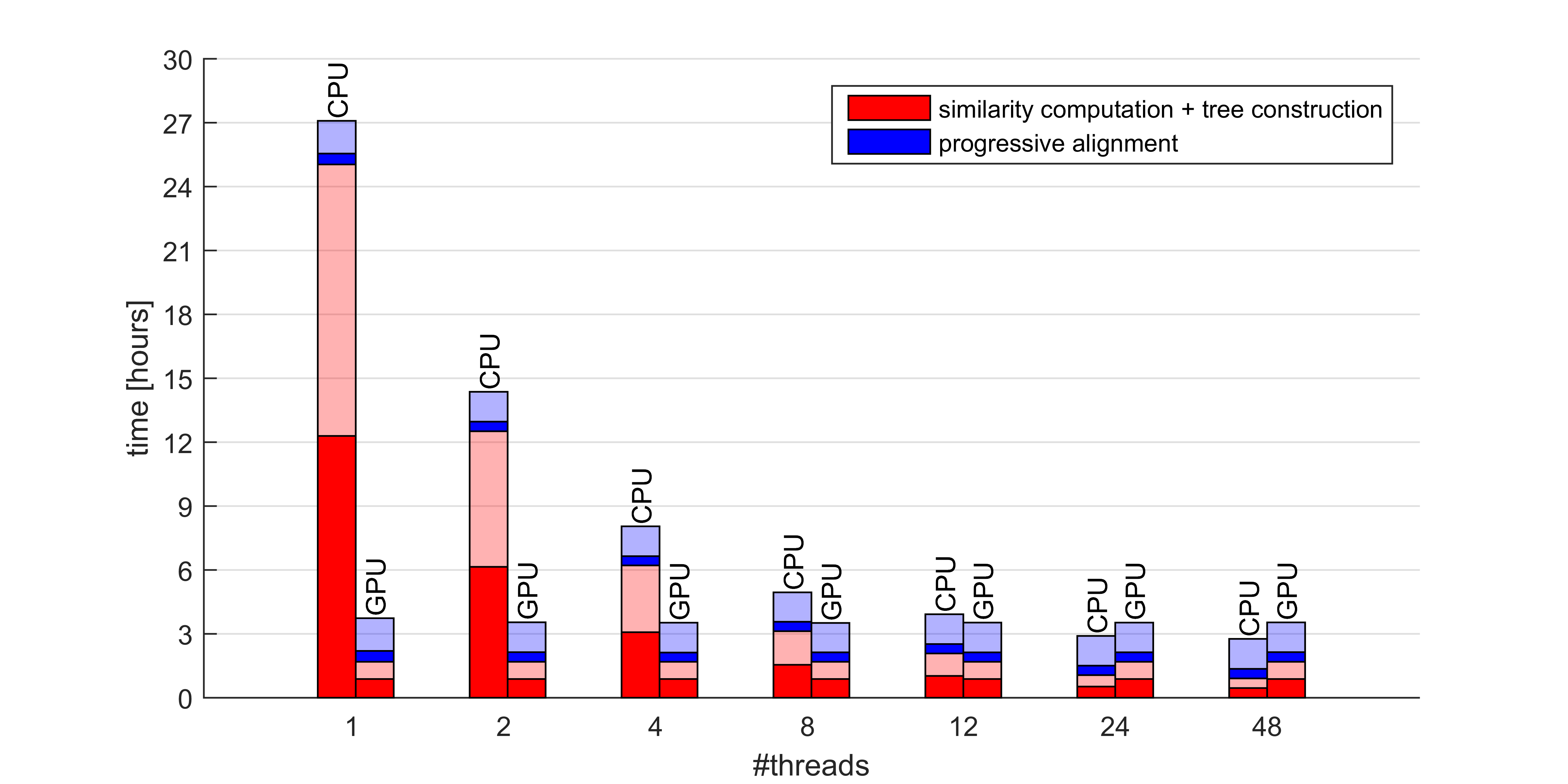}	
	\caption{Computational scalability of FAMSA with respect to the number of cores evaluated on ten largest extHomFam families ($k \geq 100\,000$). Algorithm stages are represented by different colours. Execution times of the largest set (\emph{ABC\_tran}) are marked with solid fill, the other families are printed in with transparency.}
	\label{Figure:times}
\end{figure*}

As FAMSA was designed to fully utilise available computational power, it takes advantage of multi-core architectures of nowadays computers. Ten largest protein families from extHomFam 
(all that contain at least 100\,000 sequences: \emph{ABC\_tran}, \emph{gtp}, \emph{HATPase\_c}, \emph{helicase\_NC}, \emph{kinase}, \emph{mdd}, \emph{response\_reg}, \emph{rvp}, \emph{sdr}, \emph{TyrKc})
were selected to investigate scalability of the algorithm stages with respect to the number of computing threads. In the experiments we also considered the variant of FAMSA in which similarity calculation was suited for massively parallel architectures with a use of OpenCL. For convenience, processing times of \textit{ABC\_tran} were marked separately. As Figure~\ref{Figure:times} shows, when FAMSA was run serially, more than 90\% of the execution time was related to stages I and II (the algorithm performs them simultaneously). Nevertheless, as pairwise similarities can be calculated independently, these stages scale with the number of threads noticeably better than the progressive construction. Particularly, when more than 12 cores were involved, stage III of the algorithm started to be the bottleneck. This was also the case for the GPU FAMSA variant. 

\subsection*{Classic benchmark evaluation}

For completeness, the accuracy of algorithms was investigated on classic benchmarks with families ranging from a few to approximately a~hundred of sequences (Table~\ref{tab:res:small}). According to the expectations, consistency-based methods (QuickProbs~2, MSAProbs, and GLProbs) are superior to the competitors. When non-consistency approaches are of interest, FAMSA is the second best algorithm on BAliBASE and PREFAB, and the third on OXBench-X. Interestingly, it takes the lead on SABmark.
%, which is probably caused by the presence of distantly related sequences for which XXX feature of FAMSA is especially beneficial. 
The analysis of execution times confirms FAMSA to be one of the fastest algorithms for low and moderately-sized sets as those contained in investigated benchmarks.
  
\begin{table*}[hbt]
\caption{Comparison of algorithms for small datasets.}
\small
\label{tab:res:small}
\begin{tabular*}{\hsize}{@{\extracolsep{\fill}}lcccrccrcccrcccr}\toprule
Algorithm	&&	
	\multicolumn{3}{c}{BAliBASE}&&	
	\multicolumn{2}{c}{PREFAB}&&
	\multicolumn{3}{c}{OXBench-X}&&
	\multicolumn{3}{c}{SABmark}\\ 
	\cline{3-5}	\cline{7-8}	\cline{10-12}	\cline{14-16}
			&& SP	& TC	& time	&& SP/TC	& time	&& SP	& TC	& time	&& SP	& TC	& time	\\
\midrule
QuickProbs	2	&& 88.0	& 61.7	&	23:41		&& 74.2	& 1:41:26	&& 
						89.5	& 80.3	& 1:35:35	&& 61.1	&	40.8	& 24	\\
MSAProbs			&& 87.8	& 60.8	& 35:29		&&	73.7	&	2:26:39	&&
						89.1	& 80.0	& 2:42:09	&& 60.2	& 40.0	& 29 	\\
GLProbs			&&	87.9	& 59.3	& 23:21		&& 72.4	& 1:25:40	&&
						89.1	& 80.0	& 1:10:08 	&&	61.4	& 41.4	& 3:55	\\
MAFFT auto		&&	86.5	& 58.7	& 10:20		&& 72.6	& 17:28	&&
						88.7	& 79.4	& 7:13		&& 57.3	&	36.8	& 1:01	\\
Clustal-iter2	&&	84.8	& 56.7	& 67:32		&& 71.0	& 2:35:46	&&
						88.5	& 79.5	& 45:30		&& 55.2	& 35.7	& 2:52 	\\
Clustal-default	&&	84.2	& 55.9	& 7:41		&& 70.0	& 21:56	&&
						87.8	& 78.1	& 7:34		&& 55.0	& 35.5	& 3:52 	\\
FAMSA				&&	83.6	& 53.5	& 2:01		&& 68.1	& 5:23		&&
						87.3	& 77.2	& 1:09		&& 56.9	& 37.6	&	19	\\
Kalign-LCS		&&	83.0	& 50.4	& 29			&&	65.9	&	1:51 &&
						86.8	& 76.4	& 36			&& 55.6	& 35.6	& 2	\\
Muscle			&& 81.9	& 47.8	& 14:10		&& 67.7	& 35:04	&&
						87.5	& 77.6	& 26:44 		&& 54.5	& 33.5	& 45 	\\
MAFFT default	&&	81.7	& 47.5	& 1:48		&&	68.0	& 5:58	&&
						86.6	& 76.2	& 1:39		&& 53.2	& 33.0	& 39 	\\
Kalign2			&&	81.1	& 47.1	& 37			&& 65.5	& 2:03	&&
						86.3	& 75.9	& 48			&& 52.4	& 32.6	& 2	\\
\bottomrule
\end{tabular*}
\end{table*}

\section*{Discussion}
%We introduced a new multiple sequence alignment algorithm designed especially for large protein families.
%The evaluation shows that its accuracy is superior to the existing algorithms for datasets, i.e., containing at least a few thousand of sequences.
%Moreover, it is by an order of magnitude faster than existing solutions.
%What is also interesting even the largest protein family (containing 415\,519 sequences) could be aligned on a laptop, as FAMSA required only less than 8\,GB of RAM.
%This is in contrast to the competitors as Clustal Omega failed to complete even when 128\,GB of RAM were available and MAFFT in the memory-efficient mode allocated 47\,GB.

The abundance of protein families containing hundreds of thousands members imposes development of algorithms computationally feasible to align immense sets of sequences. Traditional progressive scheme was successfully modified by Clustal Omega and MAFFT aligners to eliminate its greatest bottleneck in large-scale analyses---calculation of all pairwise similarities. Nevertheless, experiments with FAMSA show, that computation of entire similarity matrix with the use of LCS measure combined with memory-efficient single-linkage tree construction and in-place profile alignment, is orders of magnitude faster than competing solutions. Importantly, this comes with superior alignment quality---FAMSA was significantly more accurate than Clustal Omega and MAFFT on sets of a few thousands and more sequences. \emph{ABC\_tran}, the largest from investigated families with 415\,519 sequences reveals the potential of presented software. The set was processed by FAMSA within less than 2 hours in less than 8\,GB of RAM, which is suitable for a typical laptop computer. In contrast, Clustal Omega crashed after 2 days of computations on 128\,GB machine due to excessive memory requirements. MAFFT in memory-efficient mode completed the analysis in 5 hours allocating 47\,GB of RAM, yet it successfully aligned only 5.7\% of columns, while FAMSA restored 21.3\%.

The scalability of presented algorithm in terms of alignment quality as well as time and memory requirements, makes it applicable for protein families even of a million sequences---the no-go area for competing software. 
Such families will likely be present in PFam database in the near future, as a consequence of advances in sequencing technologies.
Importantly, the efficiency of FAMSA has the potential to be further increased.
The natural option is the parallelisation of the dynamic programming procedure at the profile construction stage, as it appeared to be a bottleneck in the scalability tests. Another possibility could be better utilisation of massively parallel architectures by optimising OpenCL code for GPUs or adapting it for Intel Xeon Phi co-processors.

An alternative development direction concerns alignment quality. Iterative refinement is one of numerous techniques designed for accuracy improvement. Due to computational reasons, it is performed by FAMSA on families of less than 1000 sequences, though. One can consider applying some limited, less time-consuming refinement scheme also for larger sets of sequences. The different ideas 
include introducing profile Markov models or consistency. Until recently the latter was found infeasible for large families because of excessive computational requirements. However, our latest research~\cite{GD2015} showed that applying consistency only on a small, carefully selected fraction of sequences, may elevate alignment quality without compromising execution time. The experiments concerned sets up to thousand of sequences, accordingly the scalability of presented ideas to families two orders of magnitude larger is an open question. Moreover, designing consistency scheme suitable for FAMSA is a non-trivial task. 

The separate issue related to large-scale analyses is an accuracy assessment, particularly the lack of reference sequences. Evaluating quality of alignment of 10\,000 or more proteins on the basis of a reference containing only small fraction of members is the largest flaw of the experimental pipeline used in the current research. We believe that advances in multiple alignment domain should be facilitated with the development of new benchmark datasets containing more reference sequences.   

FAMSA executables together with source code are available at \url{https://github.com/refresh-bio/FAMSA}, extHomFam can be downloaded from \url{http://dx.doi.org/10.7910/DVN/BO2SVW}. Web service
for remote analyses is under development.

%Since FAMSA is unmatched in terms of execution time, future plans should be also focused on alignment quality. One can consider introducing consistency to the FAMSA pipeline. Until recently, this technique was found infeasible for large sets of sequences due to excessive computational requirements. However, our latest research~\cite{GD2015} show that applying consistency only on a small, carefully selected fraction of sequences, may elevate alignment quality without compromising execution time. Therefore, we believe that applying consistency in large-scale analyses is a promising direction in a development of multiple sequence alignment algorithms.

\section*{Acknowledgements}
The work was supported by Polish National Science Centre under the projects DEC-2011/03/B/ST6/01588 and DEC-2015/17\-/B/ST6/01890 and by Silesian University of Technology under the project BK-263/RAu2/2015, performed using the infrastructure supported by POIG.02.03.01-24-099/13 grant: `GeCONiI---Upper Silesian Center for Computational Science and Engineering'.

\section*{Author contributions statement}
SD and ADG designed the main part of the algorithm.
AG and SD designed the GPU part of the algorithm.
SD, ADG, and AG prepared the implementation.
AG and SD designed and prepared the new benchmark.
SD performed the experiments.
SD, AG, and ADG drafted the manuscript and supplementary material.
All authors read and approved the final manuscript.

\section*{Additional information}
The supplementary material contains details on how the data were prepared and how the experiments were performed.

\subsection*{Competing financial interests}
The authors declare no competing financial interests.

%The corresponding author is responsible for submitting a \href{http://www.nature.com/srep/policies/index.html#competing}{competing financial interests statement} on behalf of all authors of the paper. This statement must be included in the submitted article file.

\clearpage

\vfill
\begin{center}
\LARGE
Supplementary material
\end{center}
\vfill
\vfill
\clearpage

\section*{Examined programs}

The following programs were used in the experimental part. 
The running parameters are also given.
\begin{itemize}
\item Clustal Omega v.\ 1.2.0
\begin{itemize}
\item \texttt{-i $<$input$>$ -o $<$output$>$ --threads=8}
\item \texttt{-i $<$input$>$ -o $<$output$>$ --threads=8 --iter=2}
\end{itemize}

\item FAMSA v.\ 1.0
\begin{itemize}
\item \texttt{-t 8 $<$input$>$ $<$output$>$}
\end{itemize}

\item GLProbs v.\ 1.0
\begin{itemize}
\item \texttt{-num\_threads $<$input$>$ -o $<$output$>$}
\end{itemize}

\item Kalign v.\ 2.04
\begin{itemize}
\item \texttt{-quiet -i $<$input$>$ -o $<$output$>$} 
\end{itemize}

\item Kalign-LCS v.\ 2.04
\begin{itemize}
\item \texttt{-quiet -b upgma -d lcs\_indel -i $<$input$>$ -o $<$output$>$}
\end{itemize}

\item MAFFT v.\ 7.221
\begin{itemize}
\item auto: \texttt{--auto --quiet --thread 8 --anysymbol $<$input$>$}
\item default: \texttt{--quiet --thread 8 --anysymbol $<$input$>$}
\item parttree: \texttt{--quiet --thread 8 --anysymbol --parttree $<$input$>$}
\item dpparttree: \texttt{--quiet --thread 8 --anysymbol --dpparttree $<$input$>$}
\end{itemize}

\item MSAProbs v.\ 0.9.7
\begin{itemize}
\item \texttt{-num\_threads $<$input$>$ -o $<$output$>$}
\end{itemize}

\item MUSCLE v.\ 3.8.31
\begin{itemize}
\item default: \texttt{-quiet -in $<$input$>$ -out $<$output$>$}
\item maxiters2: \texttt{-quiet -in $<$input$>$ -out $<$output$>$, -maxiters 2}
\end{itemize}

\item QuickProbs v.\ 2
\begin{itemize}
\item \texttt{-t 8 $<$input$>$ -o $<$output$>$}
\end{itemize}

\end{itemize}

\section*{Additional results}

\begin{sidewaystable}[ht]

\caption{Comparison of algorithms for ExtHomFam datasets. 
Times are given in hours:minutes:seconds format.
}
\label{tab:res:ExtHomFam}\small
\setlength{\tabcolsep}{0.0em}
\begin{tabular*}{\hsize}{@{\extracolsep{\fill}}lcccrcccrcccrcccrcccr}\toprule
Algorithm	&&	
	\multicolumn{3}{c}{Small}&&	
	\multicolumn{3}{c}{Medium}&&
	\multicolumn{3}{c}{Large}&&
	\multicolumn{3}{c}{Extra large}&&
	\multicolumn{3}{c}{All}\\ 
				&&	
	\multicolumn{3}{c}{$200 < k \le 4000$}&&	
	\multicolumn{3}{c}{$4000 < k \le 10000$}&&	
	\multicolumn{3}{c}{$10000 < k < \le 25000$}&&	
	\multicolumn{3}{c}{$25000 < k \le 415519$}&&
				\\
			&&
	\multicolumn{3}{c}{113 families}&&	
	\multicolumn{3}{c}{101 families}&&	
	\multicolumn{3}{c}{96 families}&&	
	\multicolumn{3}{c}{70 families}&&
	\multicolumn{3}{c}{380 families}\\ 
	\cline{3-5}	\cline{7-9}	\cline{11-13}	\cline{15-17} 	\cline{19-21}
			&& SP	& TC	& time	&& SP	& TC	& time	&& SP	& TC	& time	&& SP	& TC	& time	&& SP	& TC	& time\\
\midrule
FAMSA-opt			&&
	82.5	& 71.3	& 1:08		&&\bf 79.3	&\bf 67.1	& 8:03	&&
	\bf71.8	&\bf 55.8	& 32:35		&&\bf 63.4	&\bf 44.4	& 6:57:41	&&\bf 75.4	&\bf 61.3	& 7:39:28	\\
Clustal Omega --iter2		&&
	\bf84.7	&\bf 74.4	& 4:51:52		&& 77.7	& 65.2	& 40:12:09	&&	
	68.5	& 52.4	& 145:50:07	&& 53.3	& 35.5	& 789:02:35	&& 73.0	& 59.2	& 979:56:45	\\
Clustal Omega		&&
	\bf82.2	&\bf 70.8	& 44:39		&& 75.3	& 61.5	& 6:10:59	&&	
	66.4	& 49.4	& 22:04:32	&& 48.9	& 30.6	& 159:27:36	&& 70.2	& 55.5	& 188:27:47	\\
Kalign-LCS			&&
	77.7	& 64.1	& 29:00	&& 74.1	& 59.6	& 12:34:24	&& 66.9	& 48.5	& 153:03:14		&&
	---	& ---	& ---	&&	---	& ---	& ---	\\
Kalign2				&&
	76.8	& 63.7	& 34:41	&& 70.9	& 57.2	& 13:56:43	&& 66.2	& 49.5	& 158:42:07		&&
	---	& ---	& ---	&&	---	& ---	& ---	\\
%MUSCLE default		&&
%			& 		&			&& ---	& ---	& ---	&&	--- & --- & --- &&
%	---	& ---	& ---	&&	---	& ---	& ---	\\
MUSCLE maxiters2	&&
	70.8	& 57.1	& 53:38	&& 55.0	& 38.6	& 11:23:48	&&	40.0 & 25.7 & 80:18:23 &&
	---	& ---	& ---	&&	---	& ---	& ---	\\
MAFFT default		&&
	79.4	& 66.8	& 8:38	&& 72.4	& 58.2	& 1:23:43	&& 65.8	& 48.3	& 7:29:33	&&
	51.3	& 31.5	& 68:34:47	&& 68.9	& 53.3	& 77:36:43	\\
MAFFT parttree		&&
	74.8	& 61.7	& 8:07	&& 64.6	& 48.3	& 51:48	&& 54.9	& 38.0	& 3:16:10	&&
	45.7	& 24.5	& 25:48:33	&& 61.7	& 45.3	& 30:04:40	\\
MAFFT dpparttree	&&
	77.8	& 65.5	& 28:54	&& 69.5	& 55.3	& 3:09:57	&& 63.4 	& 45.1 & 8:48:17	&&
	54.9	& 32.8	& 37:22:16	&& 67.7	& 51.6 & 49:49:26 	\\
\bottomrule
\end{tabular*}
\end{sidewaystable}


\begin{thebibliography}{99}
	
\bibitem{Bla2010}
G Blackshields, F Sievers, W Shi, A Wilm, and D Higgins (2010) 
Sequence embedding for fast construction of guide trees for multiple sequence alignment. 
{\it Algorithm Mol Biol} 5(1):21.
	
\bibitem{BSH2014}
K Boyce, F Sievers and DG Higgins (2014)
Simple chained guide trees give high-quality protein multiple sequence alignments.
{\it Proc Nat Acad Sci USA} 111(29):10556--10561.

\bibitem{BSH2015}
K Boyce, F Sievers and DG Higgins (2015)
Reply to Tan et al.: Differences between real and simulated proteins in multiple sequence alignments.
{\it Proc Nat Acad Sci USA} 112(2):E101.

\bibitem{BSH2015b}
K Boyce, F Sievers and DG Higgins (2015)
Instability in progressive multiple sequence alignment algorithms.
{\it Algorithm Mol Biol 10} 10:26.

\bibitem{CLPPTB2006}
S Chakrabarti, CJ Lanczycki, AR Panchenko, TM Przytycka, PA Thiessen and SH Bryant (2006)
Refining multiple sequence alignments with conserved core regions.
{\it Nucleic Acids Res} 34(9):2598--2606.

\bibitem{CMCKBEN2015}
M Chatzou et al. (2015)
Multiple sequence alignment modeling: methods and applications.
{\it Brief Bioinform} doi: 10.1093/bib/bbv099.


\bibitem{DDG2014}
S Deorowicz, A Debudaj-Grabysz and A Gudy\'s (2014)
{\it Kalign-LCS---A More Accurate and Faster Variant of Kalign2 Algorithm for the Multiple Sequence Alignment Problem},
Chapter in book Man-Machine Interactions 3, Series Advances in Intelligent Systems and Computing, Springer-Verlag Berlin Heidelberg, (A Gruca, T Czach{\'o}rski, S Kozielski, Editors) 242, pp 495--502.

\bibitem{Do2005}
ChB Do, MSP Mahabhashyam, M Brudno, and S Batzoglou (2005) 
ProbCons: Probabilistic consistency-based multiple sequence alignment. 
{\it Genome Res}, 15(2):330--340.

\bibitem{E2004}
RC Edgar (2004)
MUSCLE: a multiple sequence alignment method with reduced time and space complexity.
{\it BMC Bioinformatics} 5:113.

\bibitem{E2009}
RC Edgar (2009)
Optimizing substitution matrix choice and gap parameters for sequence alignment
{\it BMC Bioinformatics} 10:396.

\bibitem{FLPSZ1951}
K Florek, J {\L}ukaszewicz, J Perkal, H Steinhaus and S Zubrzycki (1951)
Sur la liaison et la division des points d'un ensemble fini.
{\it Colloquium Mathematicae} 2:282--285.

\bibitem{G1997}
D Gusfield (1997)
Algorithms on Strings, Trees and Sequences.
Cambridge University Press.

\bibitem{GD2014}
A Gudy\'s and S Deorowicz (2014)
QuickProbs---A Fast Multiple Sequence Alignment Algorithm Designed for Graphics Processors.
{\it PLoS One} 9(7):e103051

\bibitem{GD2015}
A Gudy\'s and S Deorowicz (2015)
{\it QuickProbs 2: towards rapid construction of high-quality alignments of large protein families}.
Preprint available at: http://arxiv.org/abs/1512.07437.

\bibitem{H2004}
H Hyyr\"o (2004)
{\it Bit-parallel LCS-length computation revisited}
In Proceedings of the 15th Australian Workshop on Combinatorial Algorithms (AWOCA 2004), pp 16--27.

\bibitem{I2015}
Intel Corporation (2015)
Intel 64 and IA-32 Architectures Software Developer's Manual.
Combined Volumes: 1, 2A, 2B, 2C, 3A, 3B, 3C and 3D.\\
\url{http://www.intel.com/content/www/us/en/processors/architectures-software-developer-manuals.html}


\bibitem{Kat2007}
K Katoh and H Toh (2007) 
{PartTree}: an algorithm to build an approximate tree from a large number of unaligned sequences.
{\it Bioinformatics} 23:372--374.

\bibitem{Katoh2008}
K Katoh and H Toh (2008) 
{Recent developments in the MAFFT multiple sequence alignment program}. 
{\it Brief Bioinform} 9(4):286--298.

\bibitem{Katoh2013}
K Katoh and DM Standley (2013)
{MAFFT multiple sequence alignment software version 7: improvements in performance and usability.} 
{\it Mol Biol Evol},
30:772--780.

\bibitem{LS2005}
T Lassmann and ELL Sonnhammer (2005)
Kalign---an accurate and fast multiple sequence alignment algorithm.
{\it BMC Bioinformatics} 6:298.

\bibitem{LFS2009}
T Lassmann, O Frings and ELL Sonnhammer (2009)
Kalign2: high-performance multiple alignment of protein and nucleotide sequences allowing external features.
{\it Nucleic Acids Res} 37:858--865.

\bibitem{LSM2010}
Y Liu, B Schmidt and D.L Maskell (2010)
{MSAProbs: multiple sequence alignment based on pair hidden Markov models and partition function posterior probabilities}.
{\it Bioinformatics} 26:1958--1964.

\bibitem{Mizuguchi1998}
K Mizuguchi, CM Deane, TL Blundell, and JP Overington (1998) 
{HOMSTRAD}: a database of protein structure alignments for homologous families.
{\it Protein Sci} 7(11):2469--2471.

\bibitem{Mut1996}
R Muth and U Manber (1996) Approximate multiple string search. 
In {\it Proceedings of the 7th Annual Symposium on Combinatorial Pattern Matching}, pp 75--86.


\bibitem{Notredame2000}
C Notredame, DG Higgins, and J Heringa (2000) 
T-{C}offee: A novel method for fast and accurate multiple sequence alignment. 
{\it J Mol Biol} 302(1):205--217.

\bibitem{PH2012}
I Plyusnin and L Holm (2012)
Comprehensive comparison of graph based multiple protein sequence alignment strategies.
{\it BMC Bioinformatics} 13:64.

\bibitem{Raghava2003}
G Raghava, G Searle, P Audley, J Barber, and G Barton (2003)
{OXBench: A benchmark for evaluation of protein multiple sequence alignment accuracy.}
{\it BMC Bioinformatics}, 4(1): 47.

\bibitem{Punta2012}
M Punta et al. (2012)
The {Pfam} protein families database. 
{\it Nucleic Acids Res} 40(D1):D281--D288.

\bibitem{SN1987}
N Saitou and M Nei (1987)
The neighbor-joining method: a new method for reconstructing phylogenetic trees.
{\it Mol Biol Evol} 4(4):406--425.

\bibitem{S1973}
R Sibson (1973)
{SLINK: An optimally efficient algorithm for the single-link cluster method}.
{\it The Computer Journal} 16:30--34.

\bibitem{SDWH2013}
F Sievers, D Dinnen, A Wilm and DG Higgins (2013)
Making automated multiple alignments of very large numbers of protein sequences.
{\it Bioinformatics} 29:989--995.

\bibitem{SWD2011}
F Sievers, et al. (2011)
%Andreas Wilm2,8, David Dineen1, Toby J Gibson3, Kevin Karplus4, Weizhong Li5, Rodrigo Lopez5,
%Hamish McWilliam5, Michael Remmert6, Johannes So¨ ding6, Julie D Thompson7 and Desmond G Higgins1
{Fast, scalable generation of high-quality protein multiple sequence alignments using Clustal Omega}.
{\it Mol Syst Biol} 7:539.

\bibitem{SM1958}
R.R Sokal and C.D Michener (1958)
A statistical method for evaluating systematic relationships.
{\it Univ Kans Sci Bull} 38:1409--1438.

\bibitem{TGLGD2015}
G Tan, M Gil, AP L\:{o}ytynoja, N Goldman and C Dessimoz (2015)
Simple chained guide trees give poorer multiple sequence alignments than inferred trees in simulation and phylogenetic benchmarks.
{\it Proc Nat Acad Sci USA} 112:E99-E100.

\bibitem{Tho1994}
JD Thompson, DG Higgins, and TJ Gibson (1994) {CLUSTAL W:} improving the sensitivity of progressive multiple sequence alignment through sequence weighting, position-specific gap penalties and weight matrix choice. 
{\it Nucleic Acids Res}, 22(22):4673--4680.

\bibitem{Thompson1999}
JD Thompson, F Plewniak, O Poch (1999) {BAliBASE:} a benchmark alignment database for the evaluation of multiple alignment programs.
{\it Bioinformatics}, 15(1):87--88.

\bibitem{Walle2005}
I Walle, I Lasters, L Wyns (2005)
{SAB}mark---a benchmark for sequence alignment that covers the entire known fold space.
{\it Bioinformatics}, 21(7):1267--1268.

\bibitem{Wu1992}
S Wu and U Manber (1992) Fast text searching: allowing errors. 
{\it Communications of the {ACM}} 35(10):83--91.


\bibitem{YT2014}
K Yamada and K Tomii (2014)
Revisiting amino acid substitution matrices for identifying distantly related proteins.
{\it Bioinformatics} 30:317--325.

\bibitem{YCYZLT2015}
Y Ye, et al. (2015)
{GLProbs: Aligning Multiple Sequences Adaptively}.
{\it IEEE/ACM Trans Comput Biol Bioinf} 12:67--78.


\end{thebibliography}
\end{document}